\documentclass[10pt,letterpaper]{article}
\usepackage{opex3}
\usepackage{cite}
\begin{document}

\title{Tunable temporal gap based on simultaneous fast and slow light in electro-optic photonic crystals}

\author{Guangzhen Li, Yuping Chen, $^{*}$ Haowei Jiang, Yi'an Liu, Xiao Liu and Xianfeng Chen}

\address{State Key Laboratory of Advanced Optical Communication Systems and Networks, Department of Physics and Astronomy,
Shanghai Jiao Tong University, 800 Dongchuan Road, Shanghai 200240, China}

\email{$^*$ypchen@sjtu.edu.cn} 



\begin{abstract}
We demonstrated a tunable temporal gap based on simultaneous fast and slow light in electro-optic photonic crystals. The light experiences an anomalous dispersion  near the transmission center and a normal dispersion away from the center, where it can be accelerated and slowed down, respectively.  We also obtained the switch between fast and slow light by adjusting the external electric filed. The observed largest temporal gap is 541 ps, which is crucial in practical event operation inside the gap. The results offer a new solution for  temporal cloak.
\end{abstract}



\section{Introduction}
In resent years, temporal cloak has been widely studied both theoretically and experimentally \cite{mccall2011spacetime,fridman2012demonstration,boyd2012optical,wu2013design,lukens2013temporal,li2013effect,chremmos2014temporal,lukens2014temporal,bony2014temporal,zhou2015temporal}. It is an analogy with spatial cloak \cite{cai2007optical} from the  space-time duality associated with diffraction and dispersion \cite{kolner1994space,kolner1989temporal}.
The first experimental demonstration of temporal cloak is presented  in an optical fibre-based system \cite{fridman2012demonstration}. The split time lens uses  four-wave mixing to impart nonlinear frequency chirp on the probe beam. The dispersive element accelerates the front part of the  probe  beam and slows down its rear part,  creating a  temporal gap of 50 ps.  However, replicating the chirp discontinuity requires extremely high bandwidth  suitable for telecommunications. To overcome the restrictions, one method for comb generation by exploiting a temporal version of Talbot effect is presented  \cite{lukens2013temporal}. The continuous-wave input is converted into a broadband frequency comb, in which  no discontinuity in the chirp rate is required. Yet in both cases  the event is effectively erased from the `history' recorded by the probe field. To solve this shortcoming, a temporal cloak is reported with the new capability not only to hide optical data, but also to concurrently transmit it along another wavelength channel for subsequent readout, masking the information from one observer while directing it to another \cite{lukens2014temporal}.
Another temporal cloak based on tunable optical delay and advance is theoretically proposed by using an optical data stream as the probe light instead of continuous wave, creating a temporal gap of 39 ps \cite{zhou2015temporal}. In this system, the event occurs in the gap can be transmitted as a useful message to the receiver.  Normally, the temporal gap need to reach the lever of nanosecond, then an event such as the modulator can be practically operated in the temporal gap.

For the various approaches  to achieving temporal cloak, one of the key elements is to create a temporal gap by modulating the group velocity, which is  mostly proceeded in fibre-based system \cite{fridman2012demonstration,lukens2013temporal,mccall2011spacetime}. However, it usually need  high pump powers  to produce large changes in the intensity-dependent refractive index, where other optical processes   could limit the ability to achieve cloak. In our previous work, we  have demonstrated the modulation of group velocity by  applying an external electric field from subluminal to superluminal  in z-axis electro-optic photonic crystal (EOPC) in theory \cite{lu2010group}. Though slow light has been observed experimentally in y-axis EOPC \cite{liu2010active}, it did not provide the evidence of fast light.

In this letter, we presented a tunable temporal gap based on the simultaneous fast and slow light in y-axis EOPC. We verified the existence of fast light both in theory and experiment. By studying the relationship between phase shift ($\Phi$) and phase-mismatching ($\Delta\beta$), we found that  the light experienced an anomalous dispersion near the transmission center and a normal dispersion away from the center, where fast and slow light occurred, respectively.   We also obtained the  switch between fast and slow light by adjusting  the external electric field. The largest observed temporal gap is 541 ps, which is crucial in practical event operation.

\section{System and theoretical analysis}
\begin{figure}[htbp]
  \centering
  \includegraphics[width=11 cm]{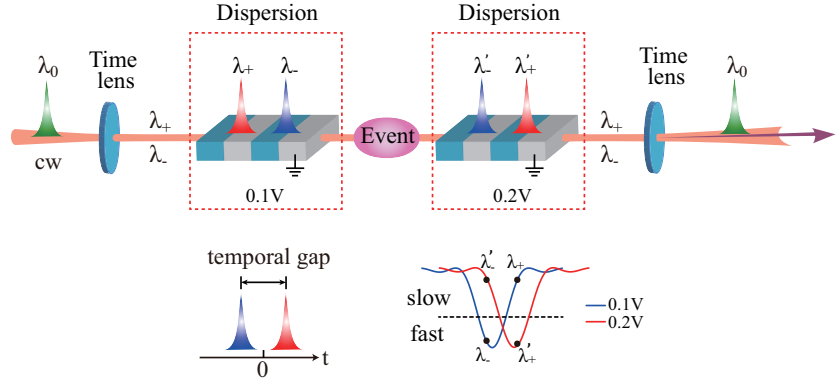}
\caption{\label{figa} Configuration of  temporal cloak based on the electro-optic photonic crystals. Time lens is used to separate or restore the continuous probe light. The EOPC acts as the dispersion element due to the second-order nonlinear effect-Pockels effect. At $E_y$=0.1 V/$\mu$m, $\lambda_+$ and $\lambda_-$  locate at the slow- and fast-light region, respectively. They are separated in time after the first EOPC, creating a temporal gap, where an event occurs. The second EOPC at $E_y$=0.2 V/$\mu$m makes $\lambda_+$ and $\lambda_-$ locate at the reverse dispersion regions and closes the temporal gap by shifting its transmission. Therefore, the event in the temporal gap remains undetected.}
\end{figure}

Figure.~\ref{figa} shows the schematic of our temporal cloak system based on the simultaneous fast and slow light. The  electro-optic photonic crystal, acting as the dispersion element, is constructed by applying y-axis external electric field on a z-cut periodically poled lithium niobate, where the optical nonlinearity can be modulated by the external electric field.
Another key element of temporal cloak is time lens, which  can produce a quadratic phase shift in time. It has been created by  electro-optic modulation \cite{kolner1989temporal,shen2010electro} and  nonlinear optics \cite{salem2008optical,ng2008compensation}. We have widely studied cascaded second-order nonlinearities  in our previous work \cite{zhang2008flexible,gong2010all,dang2013competition}. It can be used as the time lens, since we have presented that the electro-optic effects and second-order nonlinearity can be implemented simultaneously by designing the periodical structure of the nonlinear crystal \cite{li2015enhanced}.
In our configuration, the first time lens  separates the incident continuous probe wave $\lambda_0$ into two parts, $\lambda_+$ and $\lambda_-$. Then they inject to the first EOPC (with external electric field $E_y=0.1$ V/$\mu$m) simultaneously, where $\lambda_+$ and $\lambda_-$  locate at the slow- and fast-light region of the transmission, respectively. After propagating through the crystal, the two waves are separated in time since $\lambda_+$ is delayed and $\lambda_-$  is advanced, which will create a temporal gap, where an event occurs.  The gap is closed when the light passes through the second EOPC. Under $E_y$=0.2 V/$\mu$m, the transmission spectrum shifted right (about 0.3 nm in experiment, shown in Fig.~\ref{figc}(a)). The shift causes the change of phase shift, which makes  $\lambda_+$ and $\lambda_-$  locate at the reverse dispersion  regions ($\lambda_+\rightarrow\lambda_+', \lambda_-\rightarrow\lambda_-'$, and $\lambda_+=\lambda_+', \lambda_-=\lambda_-'$).
The second time lens restores the probe light to its initial state at $\lambda_0$. Therefore, the event located in the temporal gap remains undetected, which is regarded as cloaked.

Here, we give the theoretical analysis of the temporal gap. When the external electric field is applied, the principle axes of the new index ellipsoid rotate with an angle of $\pm\theta$ with respect to the unperturbed principle axes \cite{yariv1984optical,chen2003electro}. If it occurs near the phase-matching condition, the energy of the incident fundamental extraordinary wave will flow to the generated ordinary wave and then it will flow back. The returning e-polarized fundamental wave (FW) will have a different phase from the original e-wave that does not deplete completely.
The amplitudes of FWs are solved by the coupled-mode equation \cite{yariv1984optical}, which are
\begin{eqnarray}
\left\{
  \begin{array}{ll}
    A_1(z)=\displaystyle e^{i(\Delta\beta/2)z}(-i\kappa)\sin (sz)/s \\
    A_2(z)=\displaystyle e^{-i(\Delta\beta/2)z}[\cos (sz)+i\Delta\beta\sin(sz)/(2s)],
  \end{array}
\right.
\end{eqnarray}
with $\displaystyle\Delta\beta=\beta_1-\beta_2-2\pi/\Lambda$, $\kappa=-2i(n_o^{\omega}n_e^{\omega})^{3/2}\gamma_{51}E_y/\lambda$, and $s=[\kappa\kappa^{\ast}+(\Delta\beta/2)^2]^{1/2}$.
$A_1$ and $A_2$ are the normalized complex amplitudes of o-polarized and e-polarized FW; $\Delta\beta$ is the wave-vector mismatching; $\beta_1$ and $\beta_2$ are the corresponding wave vectors; $n_o^{\omega}$ and $n_e^{\omega}$ represent the indies of FWs calculated by the Sellmeier equations \cite{gayer2008temperature}; $\Lambda$ is the inversion domain period; $\kappa$ is the coupled coefficient; $\gamma_{51}$ is the electro-optic coefficient; $E_y$ is the external electric field, respectively. Then the phase shift impressed onto the incident e-polarized FW at the exit surface ($z=L$) is derived as
\begin{eqnarray}
\Delta\Phi=\frac{\Delta\beta L}{2}-\arctan[\frac{\Delta\beta}{2s}\tan(sL)].
\end{eqnarray}
And the transmission of e-polarized FW is $T=\cos^2(sz)+[\Delta\beta\sin(sz)/(2s)]^2$.

\begin{figure}[htbp]
  \centering
  \includegraphics[width=9 cm]{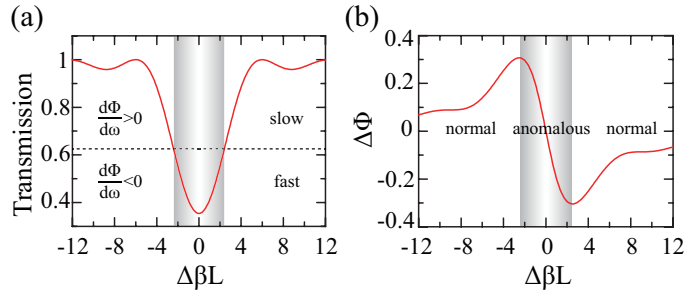}
\caption{\label{figb} (a) Transmission  spectrum and (b) phase shift of e-polarized FW as a function of $ \Delta\beta L$. The light experiences anomalous dispersion ($d\Phi/{d\omega}<0$) near the central wavelength of phase-matching (where $\Delta\beta=0$) and normal dispersion ($d\Phi/{d\omega}>0$) away from the center wavelength, where the light can be accelerated and slowed down, respectively.}
\end{figure}

Figure.~\ref{figb} shows the  transmission spectrum\cite{chen2003electro} and the phase shift as a function of $\Delta\beta L$. The polarization rotation causes the depletion of e-polarized FW, resulting in a dip in the transmission spectrum. It is similar to the dip in either a gain or absorption feature that  normally induced by electromagnetically induced transparency \cite{hau1999light} and coherent population oscillations \cite{bigelow2003observation},  where both fast and slow light can occur \cite{boyd2011material}. The light experiences an anomalous dispersion ($d\Phi/{d\omega}<0$) near the transmission center and a normal dispersion ($d\Phi/{d\omega}>0$) away from the center, where light can be accelerated and slowed down, respectively, due to the effective group velocity $1/V_g=d\Phi/d\omega/z$.  Therefore the transmission spectrum is divided into three parts. The fast-light region is near the transmission center and the two slow-light regions are near the transmission edge. The corresponding effective group index is inserted in Fig.~\ref{figd}.

\section{Experimental results and discussion}
The  electro-optic photonic crystal constructed in our experiment is  with the domain inversion period of 20.1 $\mu$m and a dimension of $40\times 10\times 0.5$ mm$^3$. The light  was launched from a continuous laser source modulated by an intensity modulator and a pulse function generator,  by which we can generate the pulse of 1 ns. The incident optical power was 10 mW.
\begin{figure}[htbp]
  \centering
  \includegraphics[width=9.5 cm]{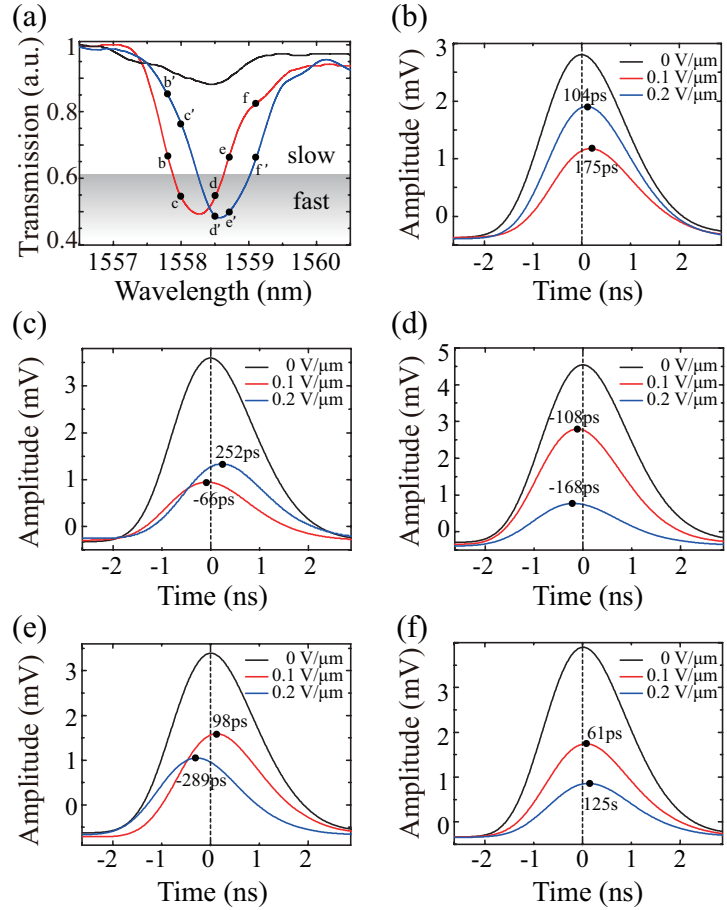}
\caption{\label{figc} Observed simultaneous fast and slow light, as well as the switch between them. (a) Measured transmission spectra under different external electric fields. The output signal waveforms with different external electric fields at fixed wavelengths of (b) 1557.8 nm (b$\rightarrow$b'), (c) 1558.0 nm (c$\rightarrow$c'), (d) 1558.5 nm (d$\rightarrow$d'), (e) 1558.7 nm (e$\rightarrow$e'), and (f) 1559.1 nm (f$\rightarrow$f'). }
\end{figure}

First, we measured the transmission spectra under different external electric fields as shown in Fig.~\ref{figc}(a). The central wavelength of the spectrum  shifted about 0.3 nm when the external electric field was changed from  0.1 V/$\mu$m to  $0.2$ V/$\mu$m.
The output signal waveforms  for different fundamental wavelengths are plotted in Figs.~\ref{figc}(b)--\ref{figc}(f), which corresponds to the  points b(b')-f(f') in Fig.~\ref{figc}(a), respectively. At 0.1 V/$\mu$m, points b, e, f are located at the slow-light region, while the other two points c and d are at the fast-light region.
When the external electric filed raised to 0.2 V/$\mu$m, the five wavelengths had quite different behaviors. At 0.2 V/$\mu$m, points b, d and f are still in their original dispersion region. The external electric field only affected their degree of delay or advance (b$\rightarrow$ b', d$\rightarrow$ d', f$\rightarrow$ f'). For point c, the pulse experienced a switch from an advancement of -66 ps to a delay of 252 ps due to the fundamental wave switching from the original fast-light region  to the slow-light region (c$\rightarrow$ c'). Similarly, point e switched from a delay of 98 ps to an advancement of -289 ps (e$\rightarrow$ e'). Then we can calculate the delay-bandwidth product of this system, which is 12.6 ( $\Delta\lambda$=0.4 nm at $\lambda$=1558 nm, $\Delta t$=252 ps).

\begin{figure}[htbp]
  \centering
  \includegraphics[width=8 cm]{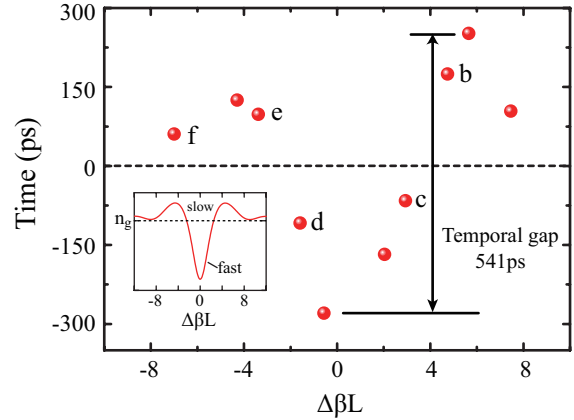}
\caption{\label{figd} Measured temporal gap as a function of $\Delta\beta L$ under $E_y$=0.1 V/$\mu$m. Points b-f correspond to the results in Fig.~\ref{figc}. The largest temporal gap is  541 ps created from -289 ps to 252 ps. The inserted figure is the theoretical effective group index as a function of $\Delta\beta L$.}
\end{figure}

To demonstrate the temporal gap, we measured the exact time for each wavelength to obtain a spectrogram of FW, which are illustrated in Fig.~\ref{figd}. The largest temporal gap is 541 ps, creating from the advancement of -289 ps to the delay of 252 ps. The observed fast- and slow-light behaviors agree well with the simulation results that inserted in  Fig.~\ref{figd}. The temporal gap can be enhanced by employing waveguide structure devices. In addition, the incident optical power is only 10 mW, which limits the occurrence of undesirable nonlinear processes.
\section{Conclusion}
In conclusion, we demonstrated a  scheme of  tunable temporal gap based on simultaneous fast and slow light in electro-optic photonic crystals.
We observed  simultaneous fast and slow light as well as the  switch between them. The temporal gap can be modulated by phase-mismatching and the external electric field.  It puts forward a new solution for temporal cloak, controlled by external electric field instead of the light intensity. It also can find other potential applications in data processing and communication security.

\section*{Acknowledgments}
The research was supported by the National Natural Science Foundation of China under Grant No.11174204, 61125503, 61235009 , and  the Participate in Research Program (PRP) of SJTU.


\begin{thebibliography}{10}
\newcommand{\enquote}[1]{``#1''}

\bibitem{mccall2011spacetime}
M.~W. McCall, A.~Favaro, P.~Kinsler, and A.~Boardman, \enquote{A spacetime
  cloak, or a history editor,} Journal of optics \textbf{13}, 024003 (2011).

\bibitem{fridman2012demonstration}
M.~Fridman, A.~Farsi, Y.~Okawachi, and A.~L. Gaeta, \enquote{Demonstration of
  temporal cloaking,} Nature \textbf{481}, 62--65 (2012).

\bibitem{boyd2012optical}
R.~W. Boyd and Z.~M. Shi, \enquote{Optical physics: How to hide in time,}
  Nature \textbf{481}, 35--36 (2012).

\bibitem{wu2013design}
K.~D. Wu and G.~P. Wang, \enquote{Design and demonstration of temporal cloaks
  with and without the time gap,} \opex~\textbf{21}, 238--244 (2013).

\bibitem{lukens2013temporal}
J.~M. Lukens, D.~E. Leaird, and A.~M. Weiner, \enquote{A temporal cloak at
  telecommunication data rate,} Nature \textbf{498}, 205--208 (2013).

\bibitem{li2013effect}
R.~B. Li, L.~Deng, E.~W. Hagley, J.~C. Bienfang, M.~G. Payne, and M.~L. Ge,
  \enquote{Effect of atomic coherence on temporal cloaking in atomic vapors,}
  \pra~\textbf{87}, 023839 (2013).

\bibitem{chremmos2014temporal}
I.~Chremmos, \enquote{Temporal cloaking with accelerating wave packets,} \ol~\textbf{39}, 4611--4614 (2014).

\bibitem{lukens2014temporal}
J.~M. Lukens, A.~J. Metcalf, D.~E. Leaird, and A.~M. Weiner, \enquote{Temporal
  cloaking for data suppression and retrieval,} Optica \textbf{1}, 372--375
  (2014).

\bibitem{bony2014temporal}
P.~Y. Bony, M.~Guasoni, P.~Morin, D.~Sugny, A.~Picozzi, H.~Jauslin, S.~Pitois,
  and J.~Fatome, \enquote{Temporal spying and concealing process in fibre-optic
  data transmission systems through polarization bypass,} Nature communications
  \textbf{5} (2014).

\bibitem{zhou2015temporal}
M.~Y. Zhou, H.~J. Liu, Q.~B. Sun, N.~Huang, and Z.~L. Wang, \enquote{Temporal
  cloak based on tunable optical delay and advance,} \opex~\textbf{23}, 6543--6553 (2015).

\bibitem{cai2007optical}
W.~Cai, U.~K. Chettiar, A.~V. Kildishev, and V.~M. Shalaev, \enquote{Optical
  cloaking with metamaterials,} Nature photonics \textbf{1}, 224--227 (2007).

\bibitem{kolner1994space}
B.~Kolner, \enquote{Space-time duality and the theory of temporal imaging,}
  Quantum Electronics, IEEE Journal of \textbf{30}, 1951--1963 (1994).

\bibitem{kolner1989temporal}
B.~H. Kolner and M.~Nazarathy, \enquote{Temporal imaging with a time lens,}
  \ol~\textbf{14}, 630--632 (1989).

\bibitem{lu2010group}
W.~J. Lu, Y.~P. Chen, X.~F. Chen, and Y.~X. Xia, \enquote{Group velocity
  modulation based on electrooptic photonic crystal with waveguide structure,}
  Photonics Technology Letters, IEEE \textbf{22}, 547--549 (2010).

\bibitem{liu2010active}
K.~Liu, W.~J. Lu, Y.~P. Chen, and X.~F. Chen, \enquote{Active control of group
  velocity by use of folded dielectric axes structures,} \apl~\textbf{97}, 071104 (2010).

\bibitem{shen2010electro}
Y.~Z. Shen, G.~L. Carr, J.~B. Murphy, T.~Y. Tsang, X.~J. Wang, and X.~Yang,
  \enquote{Electro-optic time lensing with an intense single-cycle terahertz
  pulse,} \pra~\textbf{81}, 053835 (2010).

\bibitem{salem2008optical}
R.~Salem, M.~A. Foster, A.~C. Turner, D.~F. Geraghty, M.~Lipson, and A.~L.
  Gaeta, \enquote{Optical time lens based on four-wave mixing on a silicon
  chip,} \ol~\textbf{33}, 1047--1049 (2008).

\bibitem{ng2008compensation}
T.~T. Ng, F.~Parmigiani, M.~Ibsen, Z.~Zhang, P.~Petropoulos, and D.~J.
  Richardson, \enquote{Compensation of linear distortions by using xpm with
  parabolic pulses as a time lens,} Photonics Technology Letters, IEEE
  \textbf{20}, 1097--1099 (2008).

\bibitem{zhang2008flexible}
J.~F. Zhang, Y.~P. Chen, F.~Lu, and X.~F. Chen, \enquote{Flexible wavelength
  conversion via cascaded second order nonlinearity using broadband shg in
  mgo-doped ppln,} \opex~\textbf{16}, 6957--6962 (2008).

\bibitem{gong2010all}
M.~J. Gong, Y.~P. Chen, F.~Lu, and X.~F. Chen, \enquote{All optical wavelength
  broadcast based on simultaneous type i qpm broadband sfg and shg in mgo:
  ppln,} \ol~\textbf{35}, 2672--2674 (2010).

\bibitem{dang2013competition}
W.~R. Dang, Y.~P. Chen, M.~J. Gong, and X.~F. Chen, \enquote{Competition
  between sfg and two shgs in broadband type-i qpm,} \apb~\textbf{110}, 477--482 (2013).

\bibitem{li2015enhanced}
G.~Z. Li, Y.~P. Chen, H.~W. Jiang, and X.~F. Chen, \enquote{Enhanced kerr
  electro-optic nonlinearity and its application in controlling second-harmonic
  generation,} Photon. Res.~\textbf{3}, 168--172 (2015).

\bibitem{yariv1984optical}
A.~Yariv and P.~Yeh, \emph{Optical Waves in Crystals} (Wiley, 1984), vol.~8.

\bibitem{chen2003electro}
X.~F. Chen, J.~H. Shi, Y.~P. Chen, Y.~M. Zhu, Y.~X. Xia, and Y.~L. Chen,
  \enquote{Electro-optic solc-type wavelength filter in periodically poled
  lithium niobate,} \ol~\textbf{28}, 2115--2117 (2003).

\bibitem{gayer2008temperature}
O.~Gayer, Z.~Sacks, E.~Galun, and A.~Arie, \enquote{Temperature and wavelength
  dependent refractive index equations for mgo-doped congruent and
  stoichiometric linbo3,} \apb~\textbf{91}, 343--348 (2008).

\bibitem{hau1999light}
L.~V. Hau, S.~E. Harris, Z.~Dutton, and C.~H. Behroozi, \enquote{Light speed
  reduction to 17 metres per second in an ultracold atomic gas,} Nature
  \textbf{397}, 594--598 (1999).

\bibitem{bigelow2003observation}
M.~S. Bigelow, N.~N. Lepeshkin, and R.~W. Boyd, \enquote{Observation of
  ultraslow light propagation in a ruby crystal at room temperature,} \prl~\textbf{90}, 113903 (2003).

\bibitem{boyd2011material}
R.~W. Boyd, \enquote{Material slow light and structural slow light:
  similarities and differences for nonlinear optics [invited],} \josab~\textbf{28}, A38--A44 (2011).

\end{thebibliography}
\end{document}